\documentclass[12pt]{iopart}

\usepackage{harvard}
\def\araa{ARA\&A}%
\def\apj{ApJ}%
\def\apjl{ApJ}%
\def\aap{A\&A}%
\def\aapr{A\&A~Rev.}%
\def\ssr{Space~Sci.~Rev.}%
\def\nat{Nature}%
\usepackage{graphicx}

\begin{document}

\title{Three-dimensional models of metal-poor stars}

\author{R Collet$^1$}

\address{$^1$ Max-Planck-Institut f{\"u}r Astrophysik,
Postfach 1317, D--85741 Garching b. M{\"u}nchen, 
Germany}

\ead{remo@mpa-garching.mpg.de}

\begin{abstract}
I present here the main results of recent realistic, 3D, hydrodynamical simulations of convection
at the surface of metal-poor red giant stars. 
I discuss the application of these convection simulations as time-dependent, 3D, hydrodynamical
model atmospheres to spectral line formation calculations and abundance analyses.
The impact of 3D models on derived elemental abundances is investigated by means of a 
differential comparison of the line strengths predicted in 3D under the assumption of local thermodynamic
equilibrium (LTE) with the results of analogous line formation
calculations performed with classical, 1D, hydrostatic model atmospheres.
The low surface temperatures encountered in the upper photospheric layers of 3D model atmospheres
of very metal-poor stars cause spectral lines of neutral metals and molecules to appear stronger in 3D
than in 1D calculations. Hence, 3D elemental abundances derived from such lines are 
significantly lower than estimated by analyses with 1D models.
In particular, differential 3D$-$1D LTE abundances for C, N, and O derived from CH, NH, and OH lines
are found to be in the range $-0.5$ to $-1$~dex. 
Large negative differential 3D$-$1D corrections to the Fe abundance are also computed for 
weak low-excitation Fe~{\sc i} lines.
The application of metal-poor 3D models to the spectroscopic analysis of extremely iron-poor halo stars
is discussed.

\end{abstract}

% http://www.aip.org/pacs/pacs08/pacs08-toc.html for a TOC all PACS
% http://www.aip.org/pacs/pacs08/pacs0890.html for GEOPHYSICS, ASTRONOMY, AND ASTROPHYSICS 
%hydrodynamics
%stellar atmospheres
%abundances, chemical composition
% 95.30.Tg, %thermodynamic processes, conduction, convection, equation of state

% \maketitle

\section{Introduction}
The first evidence of the existence of stars characterized by a lower abundance of metals 
with respect to hydrogen than in the Sun came in the Mid-20th century, in particular with
the works of \citeasnoun{schwarzschild1950} and \citeasnoun{roman1950}.
These authors observed that giant and dwarf stars with weaker metal lines in their
spectra typically had a larger velocity dispersion than stars with stronger spectral lines;
also, high-velocities appeared to be associated only with the weak-line stars.
These works suggested that population II stars \citeaffixed{baade1944}{as dubbed by}
differed from population I stars not only by their kinematic properties but also in terms
of basic chemical composition, the former being characterized by an overall lower iron abundance.
The following year \citeasnoun{chamberlain1951} produced the first quantitative 
abundance analysis of population II stars; the authors analysed the two stars 
HD19445 and HD140283 and derived significantly lower abundances of calcium and
iron with respect to hydrogen than in the Sun, which actually allowed to explain the peculiar
spectral and photometric properties of these objects.

It soon became clear that metal-poor stars were extremely interesting objects
which could provide important clues about the formation and the chemical and dynamical evolution
of the Galaxy.
Better tools for quantitative spectroscopy of low-metallicity stars were therefore also required, and the
first 1D model stellar atmospheres of metal-poor stars started to appear in the late 1960s-early 1970s.
Noteworthy are the works by \citeasnoun{kswamy69} and
\citeasnoun{gustafsson75}; the latter in particular constitutes the first extended grid of
theoretical 1D model atmospheres of giant stars with metallicities ranging from solar down to 
[Fe/H]$=-3$.\footnote{[Fe/H]${\equiv} \log{(n_{\mathrm{Fe}}/n_{\mathrm{H}})} - \log{(n_{\mathrm{Fe}}/n_{\mathrm{H}})}_\odot$,
where $n_{\mathrm{Fe}}$ and $n_{\mathrm{H}}$ are the number densities of Fe and H, respectively, and
the subscript $\odot$ symbol refers to the Sun.}
Since then, 1D model stellar atmospheres have been continuously developed and improved
\citeaffixed{gustafsson94,asplund05}{see, e.g., reviews by},
especially with regard to input physics and opacities, and are still nowadays the most widely used
models in stellar abundance analyses.

Yet, classical 1D model stellar atmospheres are normally constructed under a number of simplifying assumptions,
 such as plane-parallel geometry or spherical symmetry, hydrostatic equilibrium, and flux constancy.
Moreover, 1D modelling of stellar atmospheres inherently relies on rudimentary
implementations of convective energy transport, which is usually treated by means
of the mixing-length theory \cite{boehm-vitense58} or similar alternative formulations 
\citeaffixed{canuto91}{e.g}, all characterized by a number of tunable but not necessarily physical
free parameters.
In late-type stars however, the convection zone reaches and appreciably influences the surface 
layers from which the emergent stellar flux is formed. For instance, high spatial resolution
imaging of the solar photosphere immediately reveals a distinctive granulation pattern
that reflects the bulk gas motions in the upper part of the convective zone near
the optical surface.
Given the dynamic and multi-dimensional character of convection, the use of time-independent,
hydrostatic, 1D, model atmospheres in abundance analyses of late-type stars is arguably a potential 
source of severe systematic errors.
In addition, the strengths and exact shapes of spectral lines cannot possibly be predicted in 1D without
resorting to ad hoc fudge parameters such as micro- and macro-turbulence, which are supposed to 
account for non-thermal Doppler broadening due to bulk flows in the stellar atmosphere.
During the past thirty years, on the other hand, various computer codes have been 
developed to perform time-dependent, 3D, hydrodynamical simulations of convection at the
surface of late-type stars
\citeaffixed{nordlund82,nordlund90,stein98,asplund99,freytag02,ludwig02,carlsson04,voegler04}{e.g.}.
Such stellar surface convection simulations can naturally be used as 
time sequences of 3D model stellar atmospheres to study in a self-consistent way the impact of
photospheric inhomogeneities and correlated velocity fields on the formation of spectral lines
and on spectroscopic abundance analyses.
In fact, 3D simulations have been shown successful in reproducing several
observational constraints such as the morphology of the granulation pattern in the Sun as well
as detailed shapes of spectral lines in other solar-like stars.
Recent analyses based on 3D simulations of convection at the surface of the Sun, dwarfs, 
subgiants, and giants 
\citeaffixed{asplund99,asplund01,collet06,collet07,asplund05}{e.g.} indicate that
the structural differences between 3D hydrodynamical and 1D hydrostatic model atmospheres
can significantly impact the predicted strengths of spectral lines and, in turn, lead to
large differences in the derived elemental abundances, especially at low metallicities.
In the following, I will illustrate the main results of recent 3D simulations of convection at the
surface of metal-poor red giants \cite{collet06,collet07}  and discuss some applications of
these models to spectral line formation and abundance analysis.

\section{The 3D convection simulations}
The 3D, time-dependent, compressible, explicit, radiative-hydrodynamical code by  \cite{stein98} has been used
here to simulate convection at the surface of red giant stars with varying effective temperatures
($T_\mathrm{eff}{\approx}4700$ to $5100$~K), surface gravity $\log{g}=2.2$~(cgs), and 
metallicity ranging from solar down to [Fe/H]$=-3$.
The equations of mass, momentum, and energy conservation are solved on a discrete 
Cartesian mesh at moderate numerical resolution  ($100{\times}100{\times}125$) 
for a representative volume of stellar surface large enough to cover
about ten granules horizontally and eleven pressure scale heights vertically.
In terms of optical depth in the continuum at $\lambda=5000$~{\AA},  the simulations extend from
$\log{\tau_{5000}}{\approx}{-4}$ to $\log{\tau_{5000}}{\approx}7$. The upper and lower boundaries are open,
while periodical boundary conditions are assumed horizontally.
The simulations make use of realistic input physics: the equation of state comes from \citeasnoun{mihalas88},
 continuous opacities from the Uppsala opacity package 
\citeaffixed{gustafsson75}{updated version of}, line opacities from \citeasnoun{kurucz92} and \citeasnoun{kurucz93}.
The solar chemical composition by \citeasnoun{grevesse98} is adopted with
the abundances of all metals scaled proportionally to the [Fe/H] value assumed for the simulation.
The radiative heating term in the energy conservation equation is computed by solving the radiative
transfer equation at each time-step for all grid-points at the surface and along eight inclined rays plus the vertical.
To reduce the computational time, opacities are re-grouped in four \emph{opacity bins} \cite{nordlund82}.
Also, local thermodynamic equilibrium (LTE) is assumed, with a Planckian source function and
scattering treated as true absorption.

\begin{figure}
   \centering
   \resizebox{\hsize}{!}{
   \includegraphics[]{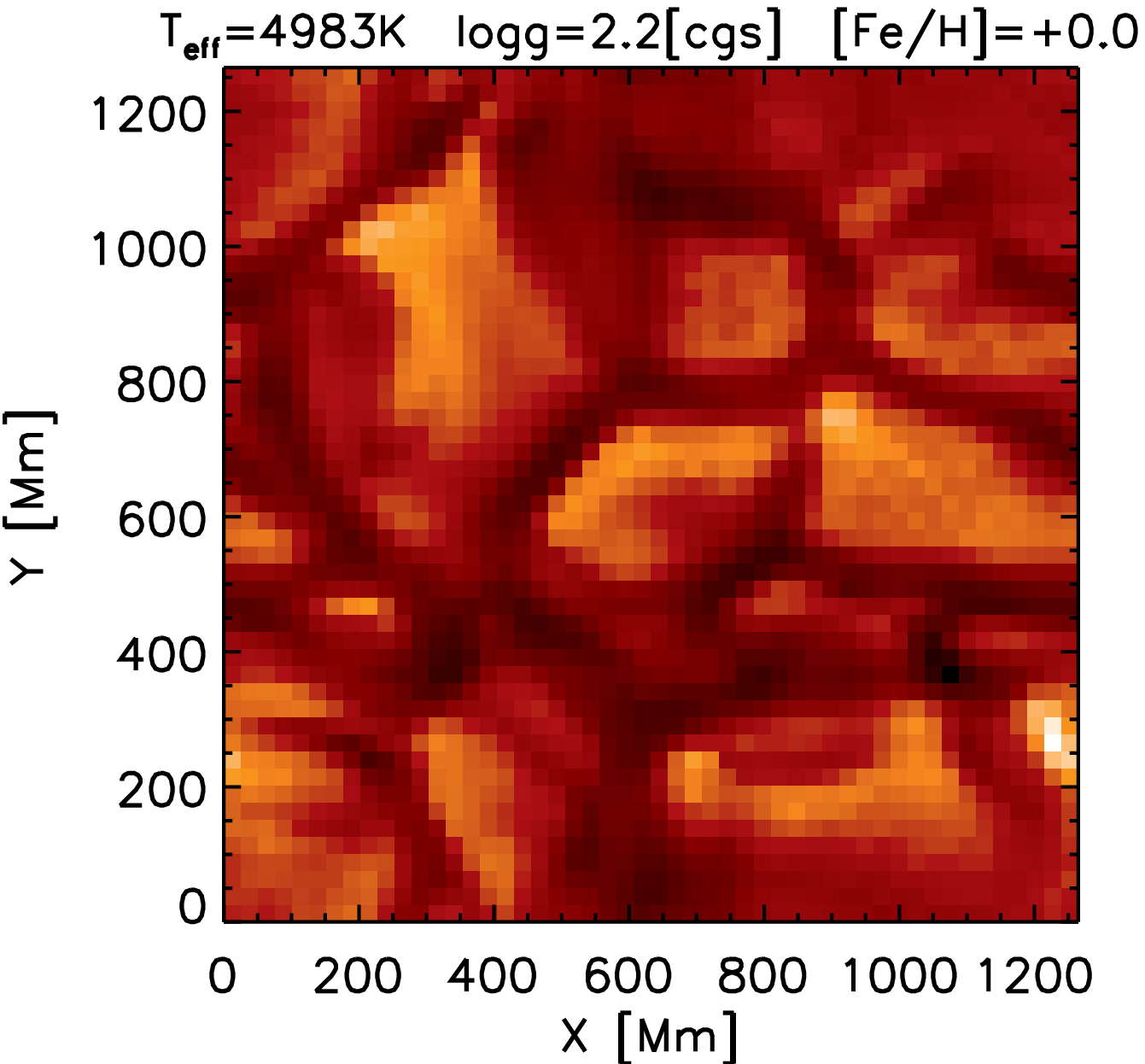}
   \includegraphics[]{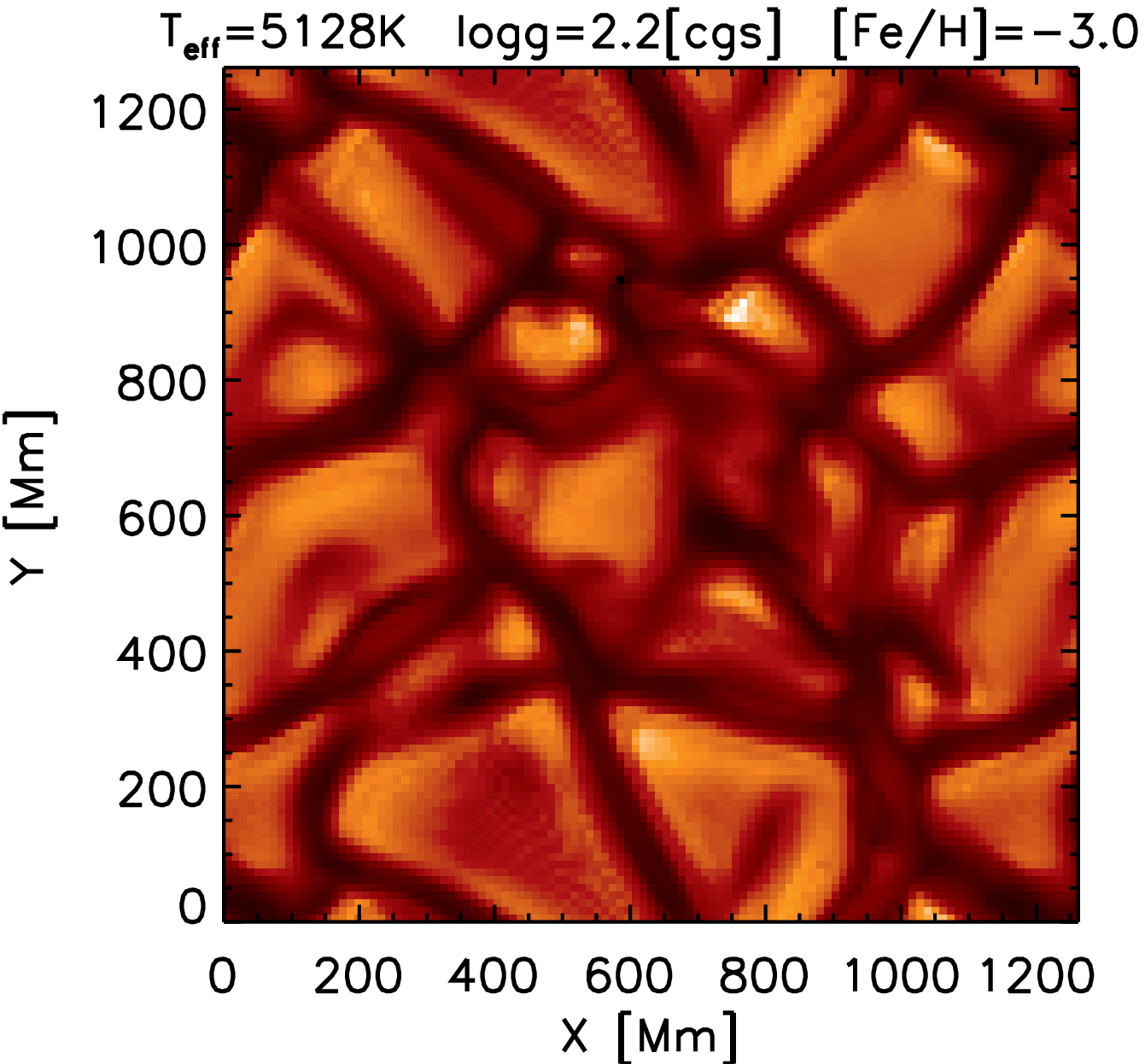} }
   \caption{Spatially resolved emergent intensity in the continuum bin for two snapshots 
   of 3D hydrodynamical simulations of red giants at  [Fe/H]$=0.0$ (\emph{left}) 
   and [Fe/H]$=-3.0$ (\emph{right});  the characteristic surface granulation pattern is apparent.}
   \label{fig:surfint}
   \end{figure}

\begin{figure}
   \centering
   \resizebox{\hsize}{!}{
   \includegraphics[]{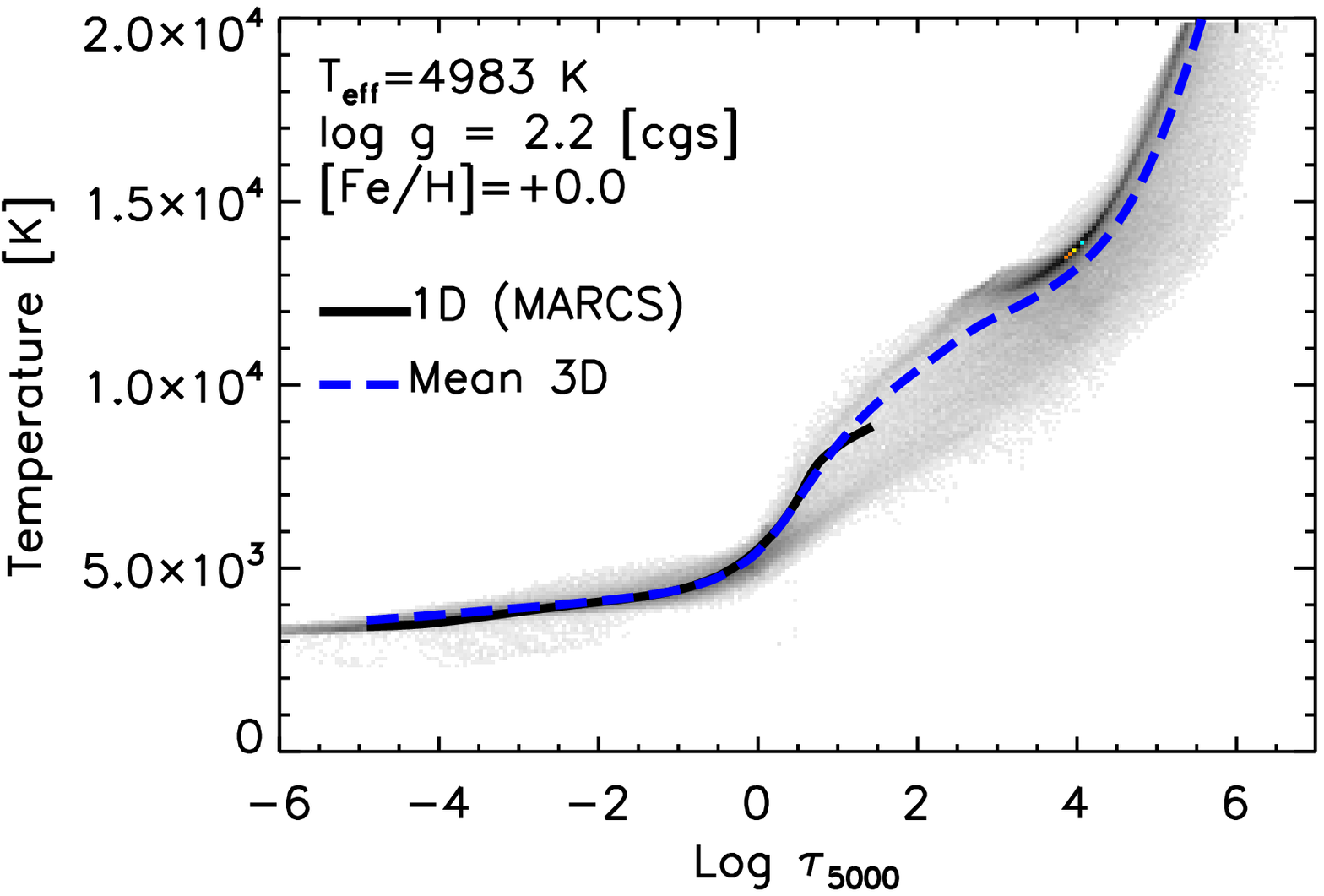}
   \includegraphics[]{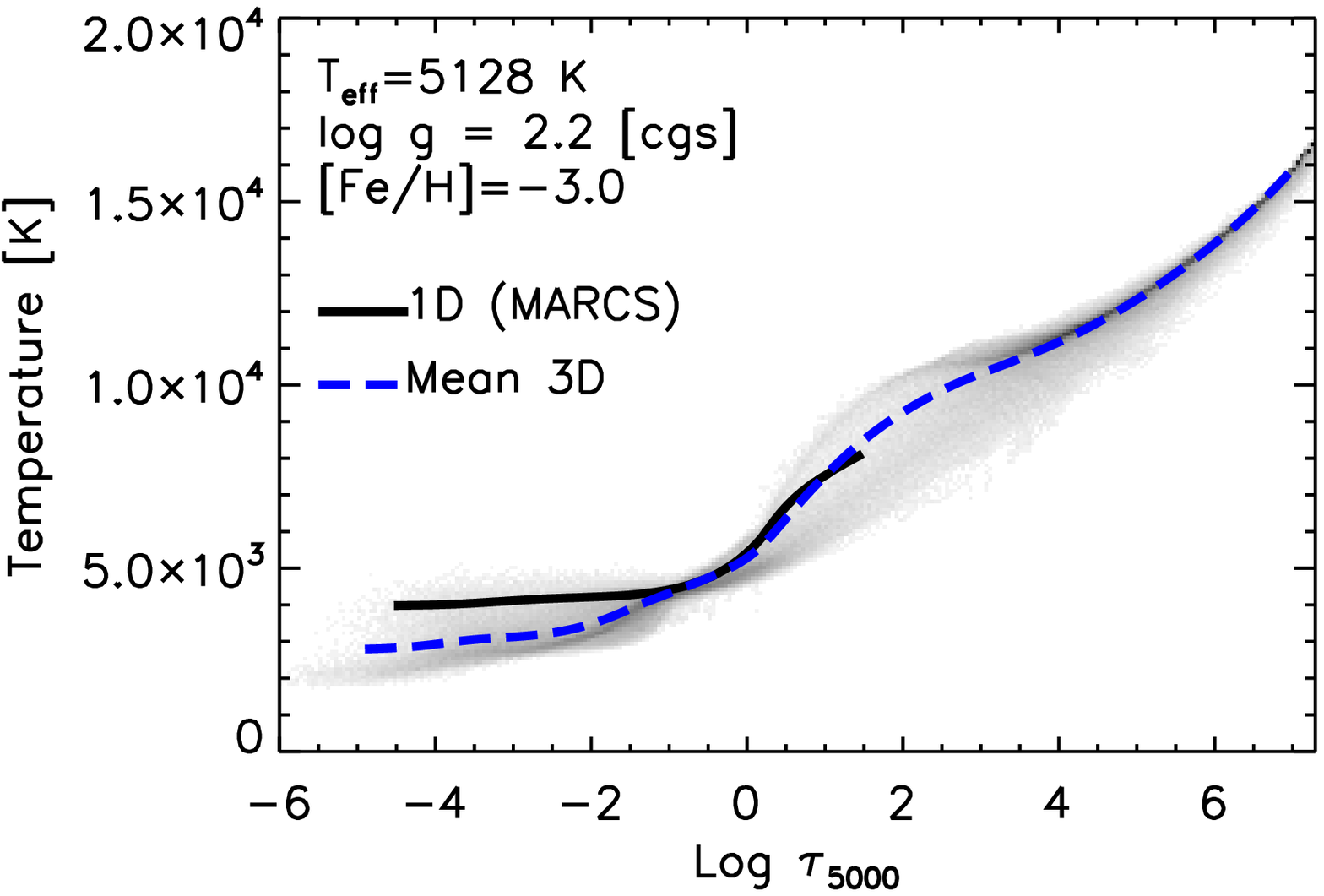} }
   \caption{Temperature structure of two snapshots of  3D hydrodynamical simulations
   of red giants at  [Fe/H]$=0.0$ (\emph{left}) and [Fe/H]$=-3.0$ (\emph{right}). 
   \emph{Gray shaded area}: temperature distribution as a function of optical depth in the continuum
   at $\lambda=5000$~{\AA}; darker areas indicate values with higher occurrence.
   \emph{Solid line}: temperature stratification of a 1D {\sc marcs} model atmosphere
   computed for the same stellar parameters.
   \emph{Dashed line}: mean temperature stratification of the 3D snapshot (temperature averages
   taken on surfaces of constant optical depth).}
   \label{fig:atmos}
   \end{figure}

The temperature and density structures at the surface of red giant convection simulations
are qualitatively fairly similar to the the ones previously described in the works by \cite{asplund99} and
\cite{asplund01} on dwarfs and turnoff stars.
Warm plasma rises from the stellar interior and rapidly cools as it approaches the optical surface; there it
eventually becomes denser than the surrounding gas, turns over, and plunges back towards the interior.
The bulk flows at the surface self-organize into a dynamical granulation pattern characterized by
large, warm upflows separated by a network of cool, narrow downdrafts.
%From a morphological point of view, the structure and evolution of the granulation in the red giant
%simulations presented here are the same as in solar-type stars.
Figure~\ref{fig:surfint} shows the spatially resolved outgoing intensity in the continuum opacity bin for
two red giant simulation snapshots with similar effective temperatures, at solar and very low ([Fe/H]$=-3$)
metallicity. The properties of the emergent granulation patterns are slightly different in the two cases: at solar metallicity,
the granules are typically larger and the overall intensity contrast is  lower than at [Fe/H]$=-3$.
More remarkable however is the metallicity dependence of the temperature stratification in the 3D simulations;
figure~\ref{fig:atmos} illustrates, as a function of optical depth, the predicted temperature structures from
the two giant simulations at [Fe/H]$=0$ and [Fe/H]$=-3$. 
The resulting thermal structures are compared with the 1D stratifications from {\sc marcs} model
atmospheres \cite{gustafsson75,asplund97} generated for the same stellar parameters, and with the same 
opacity data and chemical compositions as the 3D simulations.
At solar metallicity, the mean temperature structure in the upper photosphere of the 3D simulation appears very similar to
the stratification from the corresponding 1D model where radiative equilibrium is enforced by the stationary condition.
At very low metallicity, on the contrary, the temperature in the upper layers of the 3D simulations tends to 
remain significantly below the radiative equilibrium value marked by the corresponding 1D model.
The reason for this difference can be found by looking at the energy conservation equation.
The temperature in the upper layers of 3D simulations is mostly regulated by two competing mechanisms:
radiative heating due to the reabsorption by spectral lines of radiation coming from deeper in, and 
adiabatic cooling due to the expansion of the upflowing gas in granules.
At very low metallicity, fewer and weaker lines contribute to the total opacity, therefore the significance
of radiative heating from spectral lines is reduced, adiabatic cooling becomes preponderant,
and the balance between the two mechanisms is achieved at lower temperatures than in stationary
1D models. 

At [Fe/H]$=-3$, the temperature difference between the mean 3D and 1D photospheric
structures is substantial and can amount to ${\sim}1000$~K. 
With regard to stellar spectroscopy, such differences, as well as temperature and density inhomogeneities, 
can have a dramatic effect on the predicted strengths of spectral lines.
The cooler photospheric stratification of the 3D convection simulations of very metal-poor stars
is expected to have affect significantly temperature-sensitive features, such as molecular lines
or weak low-excitation lines from minority species (e.g. neutral metals).
Gravity-sensitive features are also affected as the lower photospheric temperatures of the 3D simulations
also imply reduced gas and electron pressures.

\section{Spectral line formation}
\label{sec:specform}
As mentioned above, 3D simulations of convection at the surface of late-type stars can be used 
as time-dependent, 3D, hydrodynamical model atmospheres for spectral line formation purposes.
In this section, I will sketch the general strategy used to compute line profiles with 3D model atmospheres
and to derive differential 3D$-$1D elemental abundances.
From the full convection simulations, representative sequences of typically ${\sim}50$-$100$ snapshots
are taken at regular intervals in time; these  time-series should cover sufficiently long periods to provide good 
statistical samples of the evolution of the surface granulation patterns.
The snapshots are interpolated to a finer depth-scale to increase the spatial resolution
in the atmospheric layers with optical depth less than $\log\tau_\mathrm{5000}{\approx}2.5$
to achieve higher numerical accuracy.
Flux profiles are computed for a variety of lines from metals and molecules.
The radiative transfer equation is solved for about $60$ to $100$~wavelength-points per spectral line profile,
along typically  ${\sim}10$-$30$~rays, and for all grid-points at the surface of the 3D model.
As during the construction of the simulation, the assumption of LTE is made, with scattering treated
as true absorption.
The same numerical code, input physics, and opacity data are used to compute spectral line profiles 
with 1D model atmospheres corresponding to the same stellar parameters; 
in the 1D calculations, a non-zero value of the micro-turbulence is adopted to mimic non-thermal 
Doppler broadening of the lines.
I would like to emphasize once more here that in 3D, on the contrary, no free parameters such
as micro- or macro-turbulence enter the calculations: only the velocity fields predicted 
by the convection simulations are taken into account to reproduce non-thermal broadening, 
asymmetries, and wavelength shifts associated with the bulk motions of the gas in the stellar
atmosphere.

The impact of 3D hydrodynamical models on stellar spectroscopy is evaluated by means of
a differential 3D$-$1D abundance analysis based on a simple \emph{curve-of-growth} method. 
The abundance of the trace element is varied independently in the 3D and 1D calculations 
to match the equivalent width of a given spectral line; the difference between the two abundances
represents the differential 3D$-$1D correction to the abundance of the trace element derived
from that particular spectral feature.

\section{Results}
\begin{figure}
  \begin{center}
  \resizebox{\hsize}{!}{\includegraphics[]{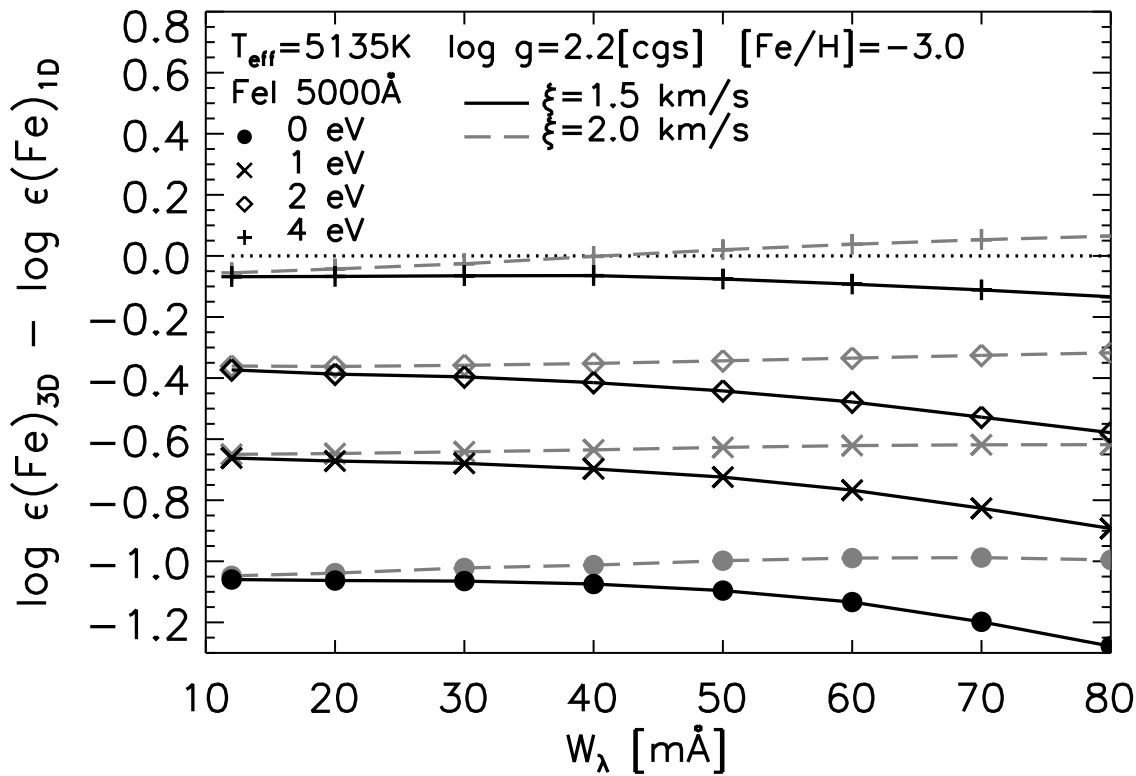} }
  \caption{Differences between the Fe abundances derived in LTE from ``fictitious'' Fe~{\sc i} lines
  at $\lambda=5000$~{\AA} with a 3D hydrodynamical and a 1D {\sc marcs} model atmosphere
  of very metal-poor red giant.
  The differential abundances are plotted as a function of equivalent width $W_\lambda$ and
  lower-level excitation potential of the lines, and for two different choices of the micro-turbulence parameter
  $\xi$ in the 1D calculations.}
  \label{fig:corr}
  \end{center}
  \end{figure}

\begin{figure}
   \centering
   \resizebox{\hsize}{!}{ \includegraphics[]{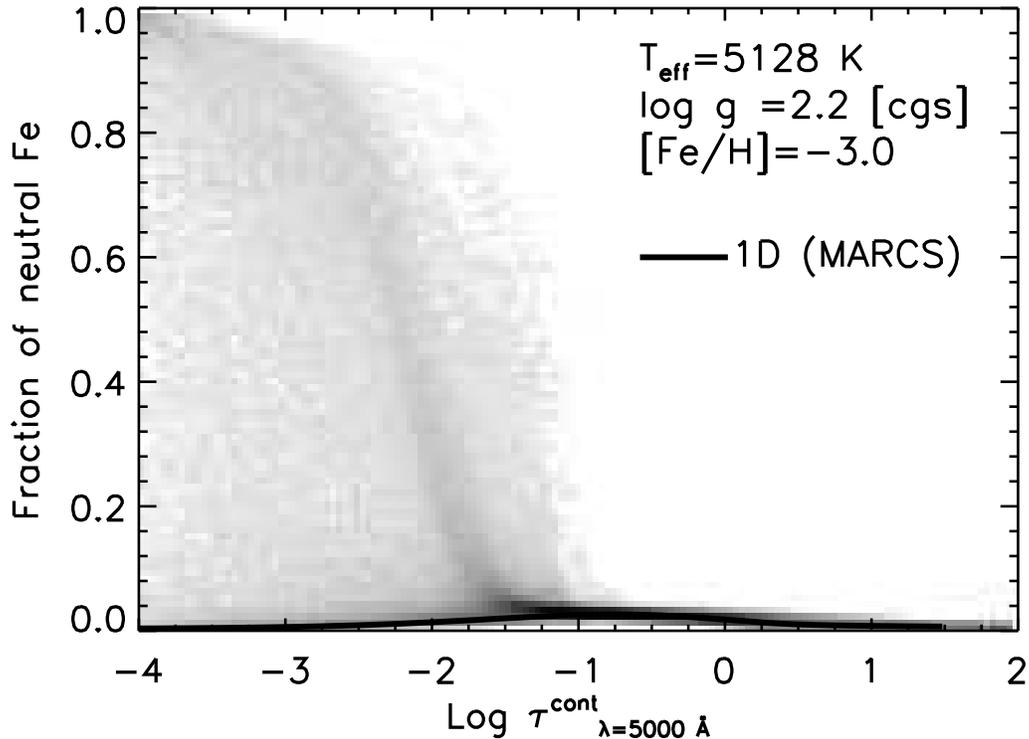} }
   \caption{Ratio of neutral to total iron number densities ($n_\mathrm{Fe I}/n_\mathrm{Fe I}$) as a function of optical 
    depth in the continuum at $5000$~{\AA} in the atmosphere of a red giant star at [Fe/H]=$-3$.
    \emph{Gray shaded area}: distribution of $n_\mathrm{Fe I}/n_\mathrm{Fe I}$ values predicted with the use of a 
    3D model atmosphere: darker areas indicate values with higher occurrence. Over-plotted is 
    the fraction of neutral iron as a function of optical depth in the corresponding 1D {\sc marcs}
     model atmosphere (\emph{solid line}). }
   \label{fig:feifrac}
   \end{figure}

%Differential 3D$-$1D LTE abundance corrections are anticipated to be
%largest for very low metallicity stars, where the predicted photospheric structure varies significantly
%depending on whether a 1D or 3D model is adopted.
%Also, such corrections are expected to be more pronounced when temperature-sensitive spectral features
%are used as abundance indicators.
In this section, I will present some representative results of differential 3D$-$1D LTE abundance
analyses at very low metallicity, where the impact of 3D models on the strengths of spectral lines 
is anticipated to be largest.
Figure~\ref{fig:corr} shows the differential 3D$-$1D LTE corrections to the Fe abundance 
for a very metal-poor red giant as derived from ``fictitious'' Fe~{\sc i} lines at $\lambda=5000$~{\AA} 
with varying equivalent widths ($W_\lambda$ between $10$ and $80$~m{\AA}) and lower-level excitation 
potentials.\footnote{Fictitious lines \cite{steffen02,asplund05,collet06,collet07} provide a
benchmark to analyse the behaviour of spectral lines solely as a function of 
lower-level excitation potential, wavelength, and strength, disregarding complications introduced 
by blends and wavelength dependency of continuous opacities in samples of real lines.}
Such corrections are overall large and negative, particularly in the case of low-excitation lines, 
for which the differences between the predicted 3D and 1D LTE Fe abundances are of the order of $-1$~dex.
The behaviour of the differential  3D$-$1D LTE iron abundances from Fe~{\sc i} lines
at low metallicity can be qualitatively explained by comparing the variations of the fraction of neutral-to-total Fe
number densities with optical depth in the 3D hydrodynamical and in the corresponding 
1D {\sc marcs} model atmospheres (Fig.~\ref{fig:feifrac}).
Under he assumption of LTE, in the 1D {\sc marcs} model, iron is nearly completely ionized 
throughout the atmosphere; on the contrary, in the 3D model, the lower surface temperatures encountered
in the upper photosphere  allow a significant fraction of iron to recombine into neutral form.
Hence, at a given Fe abundance, the density of absorbing neutral iron particles will be \emph{higher}
in the upper photosphere of the 3D model than in the 1D model and Fe~{\sc i} lines will also appear stronger.
Reversely, comparing with the 1D case, a \emph{lower} Fe abundance is required in 3D to match the same
equivalent width of a given Fe~{\sc i} line.

A similar behaviour holds for the differential 3D$-$1D LTE abundance corrections derived 
from lines of other neutral metals \cite{asplund05,collet07}.
Molecule formation as well shows an extreme sensitivity to temperature in the upper photospheric 
layers of late-type stars. This temperature sensitivity causes the 3D$-$1D LTE corrections to 
elemental abundances derived from the analysis of molecular features to also be large and
negative in very metal-poor stars \cite{asplund01,collet06,collet07}.

\section{3D abundance analysis of extreme  halo stars}
An interesting application of 3D model atmospheres of very metal-poor stars is the investigation 
of the impact of stellar granulation on the abundance analysis of the two extremely iron-poor 
halo stars HE0107$-$5240 and HE1327$-$2326 \cite{christlieb02,frebel05}.
These two stars are remarkable in the sense that the abundance of the iron-peak elements 
in their atmospheres is the lowest ever observed in stellar environments ([Fe/H]$<-5$), 
while at the same time they are characterized by very large over-abundances of carbon, 
nitrogen, and oxygen with respect to iron.
The interest aroused by these two objects comes from the consideration that  HE0107$-$5240 
and HE1327$-$2326 might be direct descendants of a previous generation of metal-free stars.
Various hypotheses have been proposed to explain the origin of the two stars but
in order to identify the most plausible formation scenario, an accurate determination
 of the their chemical composition is necessary.
\citeasnoun{collet06} and \citeasnoun{frebel08} have performed
an abundance analysis of HE0107$-$5240 and HE1327$-$2326 with the aid of 3D model 
atmospheres and following the basic procedure sketched in Sec.~\ref{sec:specform}.
The main results of the 3D$-$1D LTE abundance analysis of the two stars
are summarized in Tab.~\ref{tab:he0107}. The 3D LTE Fe abundance is estimated to be
about $-0.2$~dex lower than according to the 1D analysis, while the abundances of
carbon, nitrogen, and oxygen derived from CH, NH, and OH molecular lines
are decreased by $-0.7$~dex or more.
In the case of HE0107-5240, it is possible to derive the value of C and N abundances from more 
than one molecular indicator. Interestingly, while \citeasnoun{christlieb04} found a discrepancy 
of about $0.3$~dex between the 1D carbon abundance values derived from CH and C$_2$ lines,
the 3D analysis brings the abundances from these two indicators down to the same value
of $\log\epsilon ({\rm C}){\approx}5.7$~dex.
The 3D analysis of CN lines in HE0107-5240, however, returns an extremely large and negative
3D$-$1D correction to the nitrogen abundance (about~$-2$~dex), which is not consistent
with the 3D$-$1D correction evaluated from NH lines (see Tab.~\ref{tab:he0107}).
This discrepancy could be, on the other hand, ascribable to the highly uncertain $gf$-values
of NH lines \cite{collet06}. This problem certainly deserves further investigation.

\begin{table}
\caption{\label{tab:he0107}
Average differential 3D$-$1D LTE corrections to the C, N, and O abundances in the extremely iron-poor
halo stars HE0107$-$5240 and HE1327$-$2326 as derived from low-excitation molecular lines;
average corrections to the Fe abundance estimated from Fe~{\sc i} lines are also given.
Note: abundances are expressed in the customary logarithmic scale where $\log\epsilon(\mathrm{H})=12$.}
\begin{indented}
\lineup
\item[]\begin{tabular}{@{}*{6}{l}}
\br
 & & \centre{2}{HE0107$-$5240} & \centre{2}{HE1327$-$2326} \\
 & & \crule{2} & \crule{2}  \\
Element  & Indicator & 
$\log\epsilon_\mathrm{1D}^{\rm a}$ 	&  $\log\epsilon_\mathrm{3D}^{\rm b}$  &
$\log\epsilon_\mathrm{1D}^{\rm c}$ 		&  $\log\epsilon_\mathrm{3D}^{\rm c}$  \\
\mr
C   & CH		&  $6.81$   			& $5.75$	& $6.90$	& $6.21$ \cr
C   & C$_2$	&  $7.11$  			& $5.7$	& --		& --	        \cr 
N   & NH		&  $4.83$				& $3.8$	& $6.79$	& $6.10$ \cr  
N   & CN		&  $5.22\,(4.93)^{\rm d}$	& $3.2$	& -- 		& --	       \cr
O   & OH		&  $5.66$				& $5.0$ 	& $6.84$ 	& $6.12$ \cr
Fe & Fe~{\sc i} 	& $2.06$ 				& $1.8$ 	&  $1.79$ & $1.5$   \cr
\br
\end{tabular}
\item[] $^{\rm a}$ \cite{christlieb04,bessell04,bessell05}
\item[] $^{\rm b}$ \cite{collet06}
\item[] $^{\rm c}$ \cite{frebel08}
\item[] $^{\rm d}$ Assuming $\log\epsilon_\mathrm{1D}(C)=6.81\,(7.11)$
\end{indented}
\end{table}

\section{Discussion}
The prediction of photospheric temperatures significantly below the radiative equilibrium
value is a central result of surface convection simulations of metal-poor late-type stars. 
One implication of this result is that 3D$-$1D abundance corrections in very metal-poor stars 
are in general significantly larger than other systematic errors usually quoted in classical
abundance analyses \cite{gustafsson04b}.
It is therefore necessary to investigate to what extent the cooler temperature stratification 
of the 3D models is controlled by the assumptions made during the construction of the simulations.

\begin{figure}
  \begin{center}
  \resizebox{\hsize}{!}{\includegraphics[]{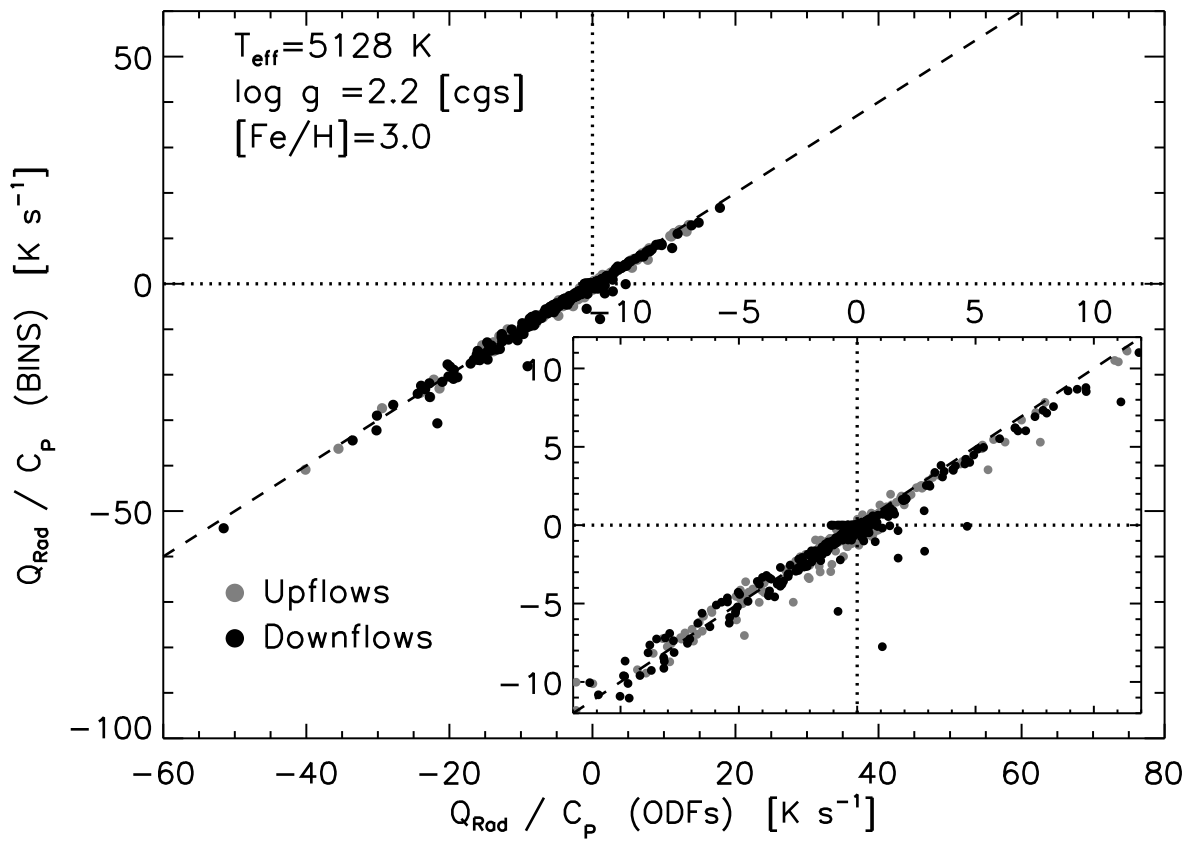} }
  \caption{Comparison between the radiative heating rates computed in ``1.5D'' approximation for all 
  grid-points on a vertical slice of a simulation snapshot of very metal-poor red giant using 
  the opacity-binning scheme (``BINS'') and monochromatic radiative transfer with opacity distribution
  functions (``ODFs'').
  The radiative heating rates per unit mass ($Q_\mathrm{rad}$) are normalized with respect to
  the speciÞc heat (per unit mass) at constant pressure ($C_\mathrm{P}$ ). 
  A magniÞed view of the plot in the region of low radiative heating rates is shown in the small panel. }
  \label{fig:qrad}
  \end{center}
  \end{figure}

A relevant question in this respect is whether the opacity-binning scheme used for the calculation of the
radiative heating rates represents a reasonably accurate approximation to the 
solution of the radiative transfer equation for the full set of wavelengths of the original
opacity distribution functions (ODFs).
Figure~\ref{fig:qrad} shows the results of test calculations where the radiative heating rates
are computed with the two different approaches in ``1.5D'' approximation along individual
 columns in a vertical slice of a metal-poor red giant simulation snapshot.
The correlation between the opacity-binning and ODF-based schemes is very good, suggesting
that the former is probably accurate enough in reproducing radiative heating and temperatures
at the surface of metal-poor 3D simulations.

   \begin{figure}
   \centering
   \resizebox{\hsize}{!}{ \includegraphics[]{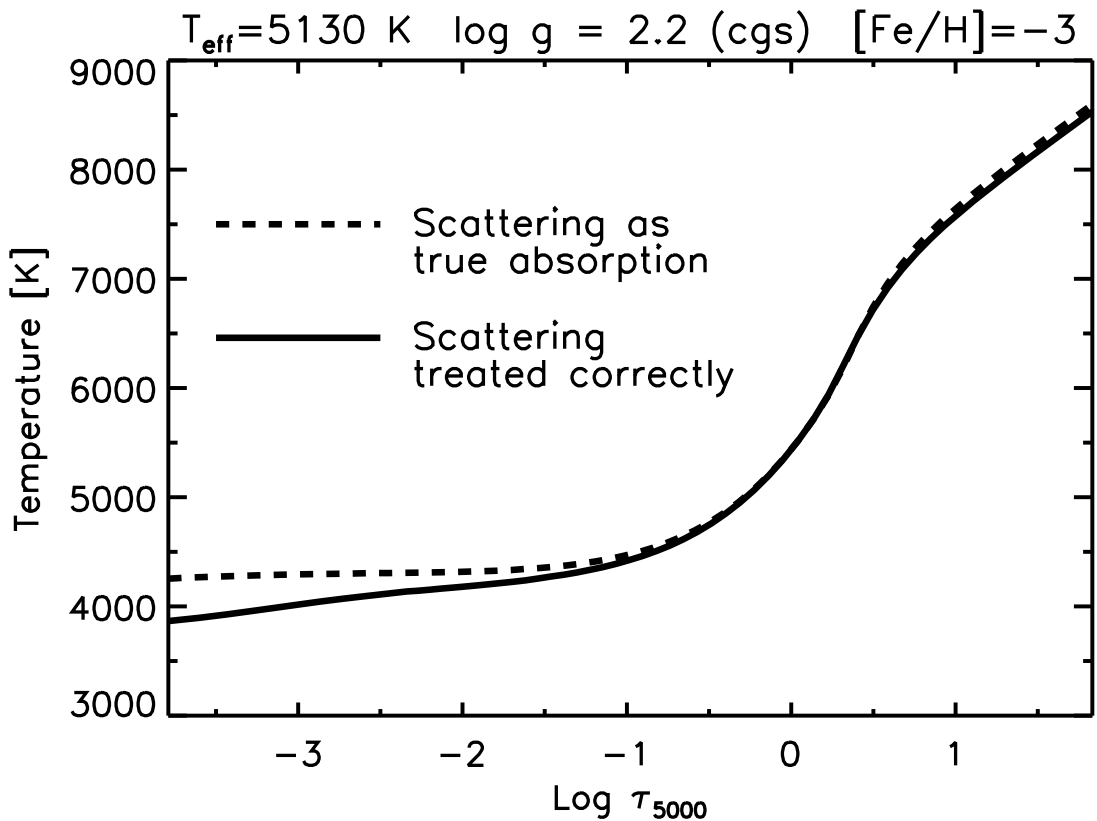}}
   \caption{Comparison between the temperature stratifications of two {\sc os-marcs}
   model metal-poor giant atmospheres computed for the same stellar parameters but with two different
   implementations of continuous scattering.  \emph{Solid line}: scattering correctly treated as such in the source function
   (standard version of {\sc os-marcs}); \emph{Dashed line}: scattering included as true absorption, as in
   the 3D simulations presented here.}
   \label{fig:scatt}
   \end{figure}

Another important issue is whether the treatment of scattering as true absorption in the solution
of the radiative transfer equation introduces systematic errors in the temperature
at the surface of 3D convection simulations.
According to preliminary tests with 1D {\sc os-marcs} \cite{gustafsson08} model atmospheres 
of metal-poor giants, the 1D temperature stratifications predicted when
scattering is included as true absorption is overall \emph{hotter} than in calculations where scattering
is treated correctly as such.
In the example considered in Fig.~\ref{fig:scatt}, temperature differences between the two cases 
reach $300$~K in the upper photosphere.
This result would therefore suggest that the radiative heating rates and temperature 
might actually be overestimated at the surface of 3D models of metal-poor stars.
It is not obvious, however, that the effect of implementing scattering as true absorption
would proceed in the same direction in 3D.
A full 3D treatment of scattering is paramount in this respect to establish whether the surface temperatures
possibly underestimated in present metal-poor 3D models. 
I refer to the contribution of W. Hayek in these proceedings for more details about 3D radiative transfer
with scattering in convection simulations.

It is also crucial to assess how spectral line formation is affected by the 
approximations involved in the solution of the radiative transfer equation.
The treatment of scattering as true absorption is a matter of concern also for 3D line formation calculations.
Rayleigh scattering off H~{\sc i} is, in fact, an important source of extinction in the UV and blue part of the spectrum.
At those wavelength, implementing scattering as true absorption causes the outgoing flux
in the continuum to be underestimated and hence leads to the prediction of \emph{weaker} spectral lines. 
The effect is expected to be more important at very low metallicities, due to the weak line-blocking, and in
in metal-poor 3D model atmospheres where the density of scatterers  (H~{\sc i}) is high because of the
low surface temperatures.
The use of a differential 3D$-$1D abundance analysis ensures at least that the uncertainties
in the treatment of scattering are minimized.

Finally, I would like to caution that many of the abundance indicators considered in the present contribution 
 (e.g. neutral metals and possibly molecules as well) are most likely affected by departures from LTE or
from chemical equilibrium at the local temperature.
Neutral iron, for instance, is expected to be prone to such departures.
The main non-LTE mechanism for Fe~{\sc i} in the photospheres of late-type stars is 
namely efficient over-ionization driven by the UV radiation field coming from layers deeper in.
This causes Fe~{\sc i} levels to be underpopulated  with respect to LTE, makes Fe~{\sc i} lines 
\emph{weaker} than the LTE case at a given Fe abundance, and, consequently, leads to the prediction 
of \emph{higher} Fe abundances.
Fe~{\sc i} departures from LTE have been estimated by \cite{collet06} for
the two extreme halo stars HE0107$-$5240 and HE1327$-$2326 by means of 
a 1D analysis (Tab.~\ref{tab:nlte}).
The results of this preliminary calculations clearly suggest that the 
non-LTE effects are considerable and opposite to the corrections for stellar granulation.
This implies that a combined 3D non-LTE analysis of  Fe~{\sc i} line formation
in a similar way as done by \citeasnoun{asplund03} and \citeasnoun{barklem03} for Li~{\sc i}
is essential to produce realistic Fe abundance determinations.
For completeness, I should also add that the departures of Fe~{\sc i} from LTE depend crucially on
the magnitude of H+Fe collisional excitation and ionization processes, whose cross-sections
are still poorly known.  This explains, at least in part, why current estimates of non-LTE
effects on Fe~{\sc i} line formation are still hampered by large uncertainties 
(see also review by A. Korn in these proceedings).

\begin{table}
\caption{ \label{tab:nlte} Non-LTE Fe abundances derived for the extreme halo stars
HE0107$-$5240 and HE1327$-$2326 using 1D {\sc marcs} models and the mean temperature
stratification versus optical depth from 3D models.
The non-LTE calculations assume the model Fe atom by \citeasnoun{collet05}, efficient
Drawin-like H$+$Fe collisions \cite{drawin68,drawin69}, and thermalization of the uppermost Fe~{\sc i} levels.}
\begin{indented}
\lineup
\item[]\begin{tabular}{@{}*{5}{l}}
\br
 &  \centre{2}{HE0107$-$5240} & \centre{2}{HE1327$-$2326} \\
 &  \crule{2} & \crule{2}  \\
Model	& [Fe/H]$_{\rm LTE}$	& [Fe/H]$_{\rm non-LTE}$ & [Fe/H]$_{\rm LTE}$ & [Fe/H]$_{\rm non-LTE}$  \\
\mr
1D {\sc marcs}	& $-5.40$ & $-4.65$ & $-5.70$ & $-5.15$    \cr
mean 3D   	& $-5.60$ & $-4.75$ & $-5.95$ & $-5.05$    \cr
\br
\end{tabular}
\end{indented}
\end{table}

\section{Conclusions}
I have presented here some illustrative results of the application of 3D surface convection simulations of 
very metal-poor stars to spectral line formation in LTE.
The differences between the predicted temperature stratifications of 3D hydrodynamical simulations 
and 1D hydrostatic model atmospheres, and the 3D temperature and density inhomogeneities and 
correlated velocity gradients can significantly affect line strengths and, in turn,
the elemental abundances inferred from spectral lines.
At very low metallicities, the deviations of the mean 3D thermal structure from the classical 1D stratification 
are largest, and cause the 3D$-$1D LTE differential abundances to be negative 
and considerable for lines of neutral species  (down to about $-1$~dex for weak low-excitation Fe~{\sc i} line).
Corrections to CNO abundances derived from weak low-excitation CH, NH, and OH features are also found 
to be typically in the range $-0.5$~dex to $-1.0$~dex in very metal-poor giants \cite{collet07}.
Finally, I have discussed possible systematic errors affecting the present 3D abundance analyses, such as
departures of Fe~{\sc i} line formation from LTE;  non-LTE corrections to the Fe abundance
are opposite to and, according to preliminary 1D test calculations, 
of the same order of magnitude as the ones due to stellar granulation.

\section*{References}
\bibliographystyle{jphysicsB}
%\bibliography{collet}

\end{document}